\documentclass[twocolumn,showpacs,amsmath,amssymb,pre,aps]{revtex4-2}   
\usepackage{graphicx}
\usepackage{dcolumn}
\usepackage{bm}
\usepackage{color}
\usepackage{amsmath}
\usepackage{lineno}
\usepackage{lastpage}
\usepackage{fancyhdr}
\definecolor{myred}{RGB}{130,0,0}
\renewcommand{\vec}[1]{\mathbf{#1}}
\newif\ifgraph

\graphtrue             %

\begin{document}
 \renewcommand{\headrulewidth}{0pt}


\title{Large-scale dynamics in visual quorum sensing chiral suspensions}
\author{Yuxin Zhou$^{1}$, Qingqing Yin$^{1}$,Shubhadip Nayak$^{2}$, Poulami Bag$^{2}$, Pulak K. Ghosh$^{2}$,Yunyun Li$^{1,*}$, Fabio Marchesoni$^{1,3}$}
\affiliation{$^{1}$ MOE Key Laboratory of Advanced Micro-Structured Materials, School of Physics Science and Engineering, Tongji University, Shanghai 200092, China}
\affiliation{$^{2}$ Department of Chemistry, Presidency University, Kolkata 700073, India}
 \affiliation{$^{3}$ Dipartimento di Fisica, Universit\`{a} di Camerino, I-62032 Camerino, Italy}
\email{yunyunli@tongji.edu.cn}
\date{\today}

\begin{abstract}
Motility-induced phase separation is an efficient aggregation mechanism of active matter, yet biological systems exhibit richer organization through communication among constituents. We investigate suspensions of active particles that change chirality when neighbor density within their visual cone exceeds a threshold - a communication-based non-reciprocal interaction akin to quorum sensing. Tuning the visual cone triggers programmable transitions—from disorder to phase separation to hyperuniformity. Notably, phase separation triggers large-scale circulation, with robust edge currents persistently flowing around dense clusters,
while particle distributions inside become effectively hyperuniform. These are genuine non-reciprocal effects which occur even in the absence of steric interactions. Remarkably, in active-passive mixtures, only $5\%$  quorum-sensing chiral particles suffice to induce collective circulation. Thus, simple perception-based rules can generate life-like order, offering design principles for programmable active materials and micro-robotic swarms.
\end{abstract}

\maketitle

\noindent
{\textcolor{myred}{\textbf{Introduction}}}\\
Our understanding of how the microscopic details of the
single-particle dynamics lead to different collective behaviors is presently
far from satisfactory. For this reason, researchers started investigating
non-reciprocal interactions as a tool to model spontaneous clustering in
active matter \cite{Granick,JP, MarchettiRev}. Non-reciprocity refers to a
scenario where a particle influences the motion of others without
experiencing any reaction in return. Non-reciprocal interactions do not
necessarily require attractive force fields \cite{Andersson,TU1,TU2} or
reciprocal pairwise potentials \cite{Golestanian}, nor do they encompass the
steric effects \cite{Fily,Redner} responsible for motility-induced phase
separation (MIPS) \cite{Cates2015}. Indeed, self-propelled particles may
aggregate by adjusting their velocity according to the perceived direction and the
local density of their peers, a mechanism known in
biology as {\em quorum sensing} (QS) \cite{QS1,QS2}.

{\em Quorum sensing} in active suspensions is known to produce intriguing
clustering effects. Bechinger and coworkers showed that disordered swarms can
form when individual particles switch off self-propulsion either in regions
of high peer concentration, independently of their relative orientation
\cite{Bauerle2018}, or, on reverse, when their sensing cone points toward low
peer density regions \cite{Lavergne2019}. We agree to refer to the latter
case as {\it visual} QS. More complex chiral QS protocols
have been invoked to generate rotating swarms, or swirls
\cite{Peruani,Bauerle2020,Israel,ourCPL,Ripoll,Gompper}.

Another unresolved question is that current MIPS models fall short in
capturing the collective dynamics of chiral swimmers \cite{NiSA}. Chirality
is an essential property of active matter. The circular and helical
trajectories executed by many biological microorganisms, like bacteria and
algae, on fluid surfaces or in chemical gradients, regulate both their
individual and collective functionalities \cite{Lauga}, like foraging and
biofilm growth. On the other hand, fabrication imperfections in the geometry,
mass density, or catalytic coating of a synthetic swimmer can result in a
(possibly unwanted) torque, which makes it a chiral active particle.
Liao and Klapp \cite{Klapp} and Ma and Ni \cite{NiJCP}  demonstrated that, when the torques are sufficiently
large, the dynamics of chiral active suspensions in two dimensions (2D) is
drastically modified: multiple clusters arise from the competition of the MIPS mechanism,
producing cohesion, and large-scale circulating currents, causing disintegration.

In this study, we explore the collective particle circulation in a chiral active
suspension by complementing (or entirely replacing) the steric pair interactions
considered in Ref. \cite{NiJCP} with long-range, non-reciprocal interactions.
To avoid the interplay with MIPS cohesion, we numerically
simulated the dynamics of strongly chiral particles with gyration radius
smaller than the average suspension mean-free path:  QS protocols were
implemented to either toggle the chirality on and off or reverse its sign
based on the local density within a limited visual cone. Focusing on the
latter protocol, we observed that, by tuning the chirality of individual particles,
visual QS can generate large-scale mass circulation in the
form of localized edge currents. Unlike the mechanism described in Ref.
\cite{NiJCP}, such currents encircle high-density particle aggregates.
As the sensing range expands, the density of these clusters decreases until they
are eventually replaced by nearly empty cavities, also surrounded by edge
currents but with an opposite sign. Ultimately, the suspension transitions to
an effective hyperuniform state \cite{NiPNAS}. Hyperuniformity (HU)
\cite{Stillinger,Torquato} characterizes the particles distribution
also within the largest clusters at lower sensing ranges. More remarkably, we noticed that, in
contrast with MIPS, such large-scale collective dynamics occurs even in the
absence of steric interactions, i.e., for pointlike particles, as a genuine
visual QS effect. When examining a mixture of passive and QS chiral
particles, we found that even a small fraction of the latter is sufficient to
produce persistent edge currents. These currents are robust enough to
``herd'' passive particles together, forming dense structures similar to
those observed in fully chiral suspensions. This finding not only reinforces
the role of edge currents in shaping two-phase structures, but also
points to potential applications in active matter manipulation and micro-robotics.

\vspace{5mm}
\noindent
{\textcolor{myred}{\textbf{Results}}}\\
\textbf{Model}\\
In 2D, the center of mass,
${\vec r}_i=(x_i,y_i)$, of an overdamped active disk of index $i$ executes a time-correlated Brownian
motion of Langevin equations (see Materials and Methods)
\begin{eqnarray} \label{LE}
\dot {\vec r} _i=  {\vec v}_{0i}, \;\; \dot \theta_i = \omega_i + \sqrt{D_\theta}~\xi_{\theta i} (t).
\end{eqnarray}
Here, the modulus of its self-propulsion vector, ${\vec v}_{0i}=v_0(\cos
\theta_i, \sin \theta_i)$, is constant, while its orientation with respect to the longitudinal $x$-axis, $\theta_i$,
fluctuates subjected to the stationary, delta-correlated noise source
$\xi_{\theta i} (t)$, with $\langle \xi_{\theta i}(t)\xi_{\theta j}(0)\rangle = 2 \delta_{ij}\delta
(t)$. The particle chiral frequency, $\omega_i$, is allowed to switch between two
fixed values according to the QS protocol detailed below.

We numerically simulated a chiral suspension by randomly placing $N$ identical,
independent Janus disks of Eq. (\ref{LE}) in a square box of side $L$ and
imposing periodic boundary conditions. Effects due to the particle finite
size were modeled by the Weeks-Chandler-Andersen (WCA) pair potential \cite{CWA},
$V_{ij} = 4\epsilon [({\sigma}/{r_{ij}})^{12} -({\sigma}/{r_{ij}})^{6}+1/4]$,
if $r_{ij} \leq r_m$, and $=0$, otherwise, where  $r_m=2^{1/6}\sigma$,
$\epsilon=1$, $i,j=1, \dots N$ are the pair labels, and $\sigma = 2r_0$
was taken as the ``nominal'' disk diameter. The average packing fraction
of the suspension is $\bar \phi=\pi r_0^2 \rho_0$, with $\rho_0=N/L^2$
denoting its average density. Further reciprocal interactions
and hydrodynamical interactions \cite{PNAS,Takagi}, have been neglected.

\begin{figure*}[ht]
\centering \includegraphics[width=18cm]{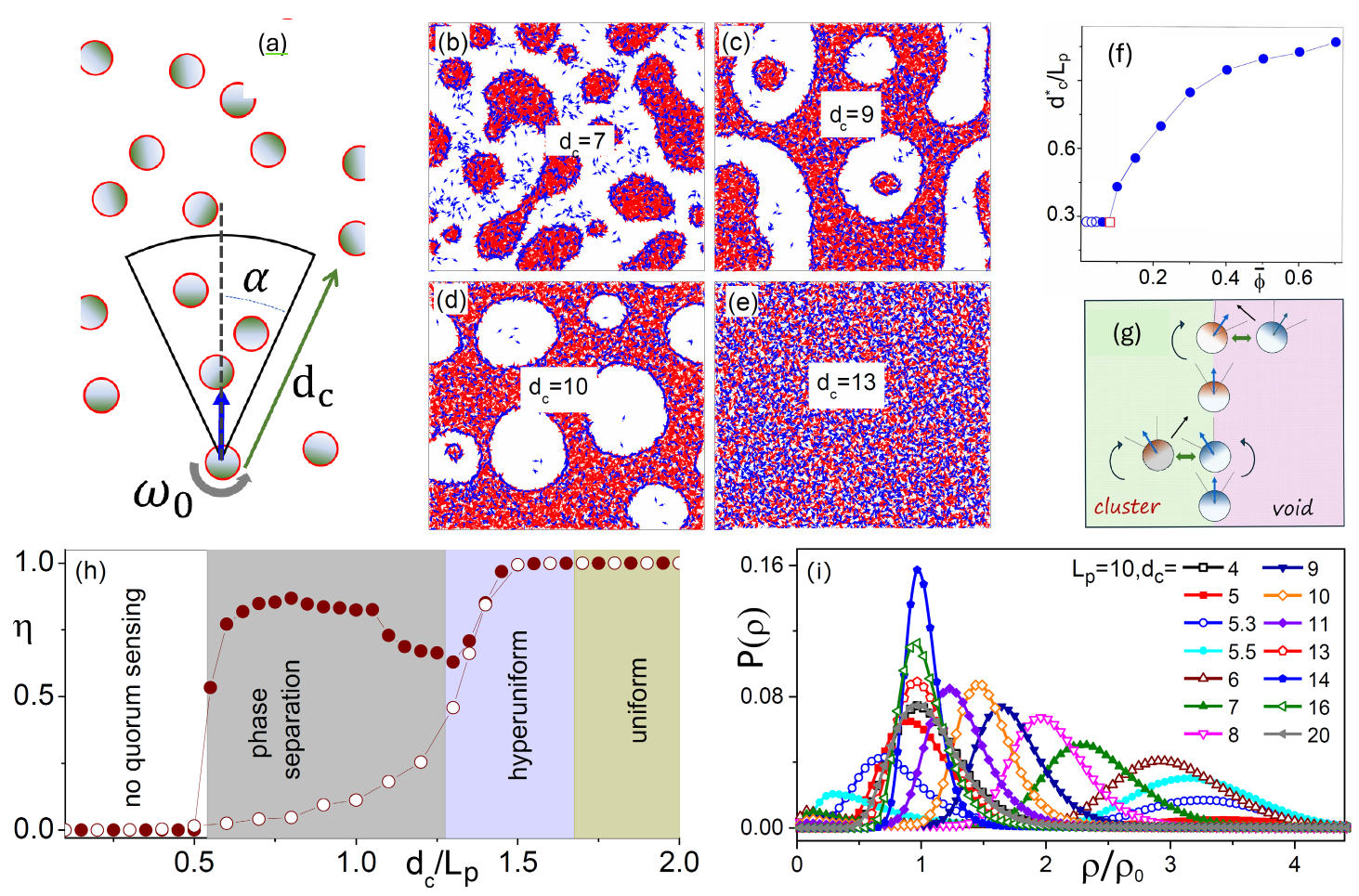}
\caption{Configurations of a chiral suspension of $N=10^4$ Janus hard disks of radius $r_0=1$, speed $v_0=0.5$ and persistence time $\tau_\theta=100$; the simulation box size, $L$, corresponds to the average packing fraction $\bar \phi=0.15$. (a) Particle's QS visual cone, $V_i^{d_c}$, introduced in Eq. (\ref{Pa}). (b)-(e) Suspension snapshots at $t=3\cdot 10^4$ for QS protocol 1 with $L_p=10$, $\alpha=\pi/2$, $\omega_0=1$, and different $d_c$ (see legends) (see also Fig. S6). Arrows of different colors denote the orientation of the self-propulsion vectors, ${\vec v}_{0i}$, of particles with positive (blue) and negative (red) chirality. (f) Clustering threshold $d_c^*$  vs. $\bar \phi$ for the model parameters of (b)-(e): filled circles correspond to the appearance of the $\rho>\rho_0$ peak in the $P(\rho)$ curves in (i); the red square denotes the critical point $(\phi_{\rm QS}, d_{\rm QS})$ introduced in the text; empty circles mark the appearance of chirality switches in the absence of clusters.  (g) Schematic of the QS-induced edge current mechanism (see text).
(h) Fraction of particles with chirality $\omega_-$, $\eta$, vs. $d_c$ in a suspension with $N=10^4$ and $\bar \phi=0.15$. $\eta$ was computed starting at $t>5\cdot 10^4$ and averaged 250 times at intervals of 200 time units under protocol 1 (filled circles) and 2 (empty circles). (i) Distribution of the local suspension density, $\rho$, for different $d_c$ (see legend) computed by means of the Voronoi polygon method \cite{Voronoi} and time averaged over 500 snapshots taken 200 time units apart, starting at $t=10^5$. Note that the higher-density peak (dense phase) emerges abruptly at $d_c \simeq 5.3$, and shifts toward $\rho_0$ as $d_c$ increases. In contrast, the lower-density peak (dilute phase) shifts toward 0, until it disappears entirely at $d_c \simeq 13$. \label{F1}}
\end{figure*}

We then assumed that the chirality of the
tagged particle depends on the spatial distribution of its neighbors.
Without entering the details of specific sensing mechanisms \cite{QS1,QS2,fish1,fish2},
we defined the sensing function of particle $i$ as \cite{Bauerle2018}
\begin{eqnarray}
\label{Pa} P_i(d_c)= \sum_{j\in V_i^{d_c}}{1}/{(2\pi r_{ij})},
\end{eqnarray}
where $V_i^{d_c}$ is its ``visual cone'' with axis parallel to ${\vec v}_{0i}$, radius (or sensing range) $d_c$, and
semi-aperture $\alpha$ [Fig. \ref{F1}(a)]. The particle chirality is then governed by the simple
{\em QS protocol},

\begin{eqnarray}
\label{QS-1}
{\omega}_i=
\begin{cases}
\omega_+ & P_i(d_c) \leq P_{\rm th} \\
\omega_- & P_i(d_c) > P_{\rm th}
\end{cases}
\end{eqnarray}
with QS threshold $P_{\rm th}= (\alpha/\pi) \rho_0 L_p$ \cite{NSO}. The focus of this report is on protocol 1:
$\omega_+=-\omega_-=\omega_0$, with tunable $\omega_0\geq 0$. Numerical
results for protocol 2: $\omega_+=0$ and $\omega_-=-\omega_0$, are
reported for a comparison (see Sec. S1 of Supplementary Text). As shown in Ref.
\cite{Ripoll}, $\omega_i$ can be regulated by tuning the misalignment
between ${\vec v}_{0i}$ and the particle's visual cone axis. Of course,
reversing the sign of $\omega_0$ does not change the qualitative description
of the suspension collective dynamics. Note that for a uniform suspension the
sensing function of Eq. (\ref{Pa}) is approximatively $\bar P(d_c)=
(\alpha/\pi)\rho_0 (d_c-\sigma)$. Clearly, this form of particle
interaction is non-reciprocal.

\begin{figure*}[ht]
\centering \includegraphics[width=20cm]{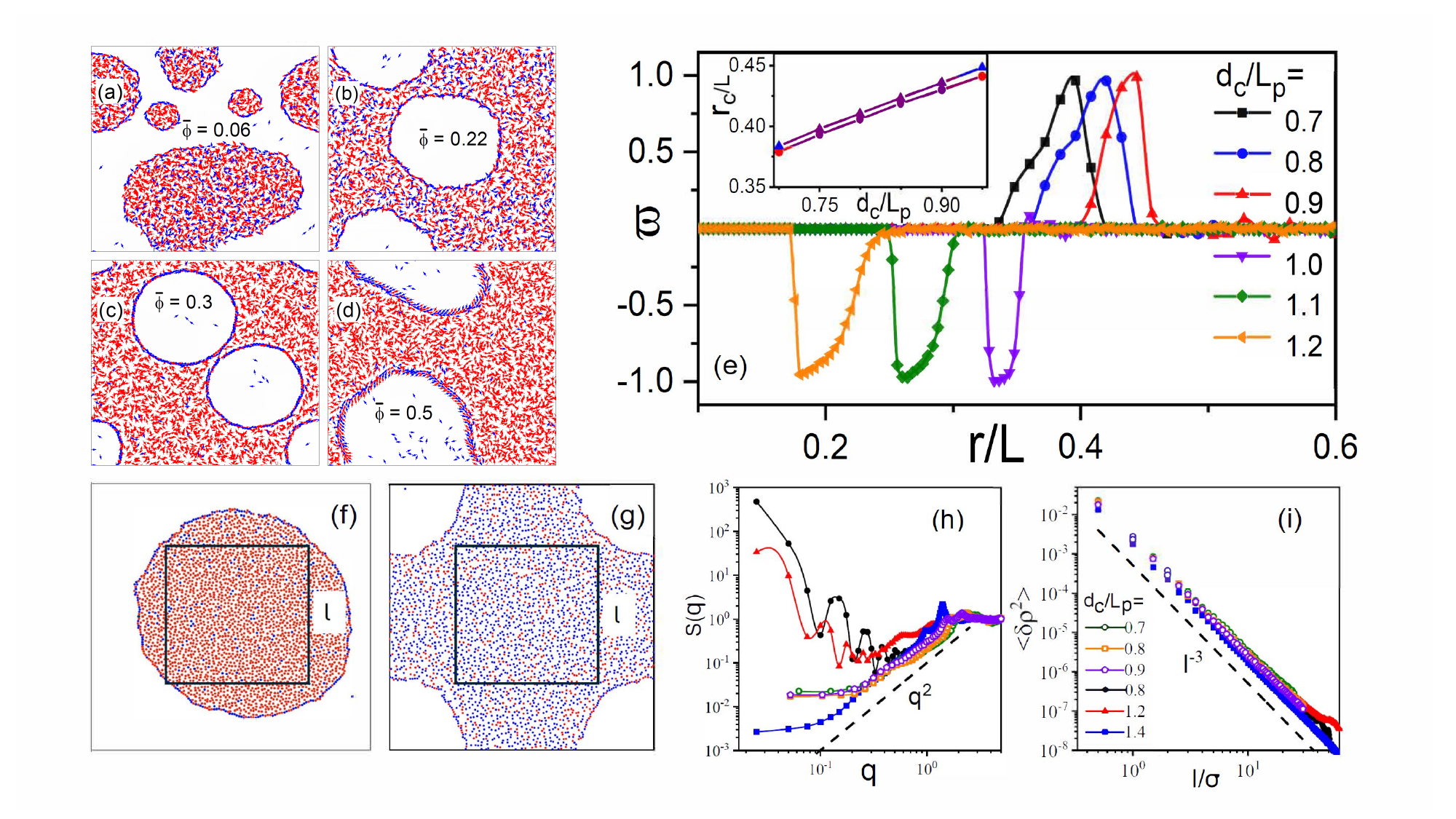}
\caption{Edge currents in two-phase configurations of a chiral suspension of $N=3\cdot10^3$ Janus disks under QS protocol 1.
(a)-(d) suspension snapshots at $t=3\cdot 10^4$ for different $\bar \phi$ (see legends); arrow color code as in Fig. \ref{F1}.
Disk parameters: $r_0=1$, $v_0=0.5$, and $\tau_\theta=100$; QS protocol parameters: $d_c=8$, $L_p=10$, $\alpha=\pi/2$, and $\omega_0=1$.
For better visualization, refer to the movies in Supplementary Material.
(e) Vorticity, $\varpi$ vs $r$, for $\bar \phi=0.15$ [positive for clusters ($d_c=7, 8, 9$) and negative for cavities ($d_c=10, 11, 12$)], time averaged over 500 snapshots taken 200 time units apart, starting at $t=10^6$. All other parameters are as in (a)-(d); $r$ is measured from the cluster (cavity) centroid. Inset: maximum of $\varpi(r)$, $r_c$, vs $d_c$ for $N=3\cdot10^3$ (blue symbols) and $N=6\cdot10^3$(red symbols).
(f)-(i) Hyperuniformity in the above chiral suspension with $N=3\cdot10^3$ and $\bar \phi=0.15$ under QS protocol 1.
(f),(g) Stationary configurations respectively for $d_c=8$ (cluster) and $10$ (cavity). The black squares denote the area where the HU quantifiers in (h) and (i) were computed (we remind for a comparison that here $L=250$). (h) $S(q)$ and (i) $\langle \delta \rho^2 (l)\rangle$ for different $d_c$ in dense phase regions (empty symbols) and over the entire simulation box (filled symbols). The suppression of the $S(q)$ decay at very low $q$ is a noise effect (see Sec. S4.F of Supplementary Text).
Time averages: over 1,000 snapshots taken 200 time units apart, starting at $t=5\cdot 10^4$.
\label{F2}}
\end{figure*}

\vspace{5mm}
\noindent
\textbf{Monodisperse suspensions}\\
We started simulating a uniformly distributed
active suspension with $\omega_i=\omega_+$, and packing fraction, $\bar \phi$,
small enough to rule out conventional MIPS \cite{Redner1,PRR7}.
We set $\omega_0$ large enough for the particle gyration
radius, $r_\omega=v_0/\omega_0$, to be much smaller than
the mean free path in a uniformly distributed achiral suspension, $l_c =\pi r_0/2\bar \phi$.
As detailed in Materials and Methods, under this assumption, the ratio $d_c/L_p$, which characterizes
the range and strength of the QS interaction, is the only relevant control parameter involving the length scales in the model.
In the absence of additional long-range interactions, the suspension distribution
remains randomly uniform \cite{NiSA}. On applying QS protocol 1, its
collective dynamics changes dramatically (Figs. \ref{F1} and \ref{F2}). On
increasing the sensing range beyond a certain threshold, $d_c^*$ [its dependence
on $\bar \phi$ is illustrated in Fig. \ref{F1}(f)], the particle distribution
grows inhomogeneous, with the formation first of multi-cluster patterns and,
then, of cavities. Contrary to conventional MIPS, with increasing $d_c$
the stationary particle density in the regions surrounding
the cluster structures and inside the cavities tends to
vanish, whereas the density inside the clusters decreases, i.e., clusters grow in size [Figs.
\ref{F1}(b)-(e), see also Fig. \ref{F2}(a)-(d)]. Accordingly, the transition from
cluster to cavity configurations inside a finite simulation box is marked by
extended periodic structures [Fig. \ref{F1}(c)] (see also Sec. S3 of Supplementary Text),
which appear for $d_c$ values independent of the system size. On
further increasing $d_c$, cavities shrink until they vanish and the
suspension turns hyperuniform \cite{NiPNAS}. Furthermore, we remark that the
transient time for the suspension to achieve its steady state diverges with
its size, $N$ (see discussion of Fig. S9).
The dependence of the configuration of a finite-size suspension
on the sensing range under protocol 1 and 2 is illustrated in Fig. \ref{F1}(h).
There we used the fraction of particles that underwent chirality change, $\eta$,
as a quantitative indicator. Under protocol 2,
no phase separation occurs and HU emerges after about 50\% of the particles
have turned chiral. In stark contrast with the uniform configurations from both
protocols, particle diffusion in the presence of two-phase structures is
mostly anomalous (see Sec. S2 of Supplementary Text).

From a more quantitative viewpoint, we remark that for the QS protocols of
Eq. (\ref{QS-1}) to describe a collective mechanism involving three or more
JP's, the sensing function of Eq. (\ref{Pa}) for a colliding pair,
$P_p(\sigma)=1/(2\pi \sigma)$, must be smaller than $P_{\rm th}$, that is
$\bar \phi>\phi_{\rm QS}$ with $\phi_{\rm QS}=(\pi/4\alpha)(r_0/L_p)$.
Moreover, since the disk diameter is finite, the further condition,
$d_c>d_{\rm QS}=\sigma$, applies, consistently with the data displayed in
Fig. \ref{F1}(f). The formation of clusters for $d_c>d_c^*$ is marked by the
appearance of a second peak in the $P(\rho)$  curves of Fig. \ref{F1}(i),
centered around a local density value, $\rho_c$, larger than $\rho_0$. Note
that $\eta$ may be positive even for $\bar \phi<\phi_{\rm QS}$, i.e.,
particles may switch chirality without forming clusters. Particles inside
large clusters [$0.6<d_c/L_p<1$ in Fig. \ref{F1}(h)], look uniformly
distributed, so that $\rho_c$ can be estimated by equating $\bar P(d_c)$ with
$P_{\rm th}$, hence $\rho_c = \rho_0 L_p/(d_c-\sigma)$, in fairly good
agreement with the simulation data, e.g., of Fig. \ref{F2}(e). This means
that here, contrary to MIPS, the density of the dense phase does not
remain constant at varying the control parameter, $d_c$. This argument is less accurate for
small clusters, where a relatively larger fraction of particles appears
to move along the boundaries in dense files. By the same token we anticipate that for $d_c
-\sigma \gtrsim L_p$  local particle density inhomogeneities are no longer
required for the tagged particle to satisfy the QS condition  Eq.
(\ref{QS-1}). This yields a working estimate for the $d_c$ value where
two-phase configurations disappear and HU sets in ($d_c/L_p \gtrsim 1.3$ for
the suspension of Fig. \ref{F1}). Based on this argument, the transition to
HU is expected to be independent of the simulation box' size.

\vspace{5mm}
\noindent
\textbf{Edge currents and hyperuniformity}\\
As anticipated above, cluster and cavity structures are
delimited by edge currents, which, for $\omega_0>0$, flow counterclockwise
(clockwise) along the cluster (cavity) boundaries. In stationary configurations, when
almost all particles aggregated into large clusters or the cavities are
empty, boundary particle circulation involves one or more layers, their
number increasing with $\bar \phi$ [Figs. \ref{F2}(a)-(d)]. This happens only for
QS protocol 1, as sketched in Fig. \ref{F1}(g). Suppose that a disk moves
along the divide between a dense region on the left and an empty one on the
right. If the peer concentration inside its visual cone is relatively low, it
will rotate counterclockwise, until it point inside the dense region. At this
point, according to QS protocol 1, it reverses its chirality sign and
starts rotating clockwise. As its visual cone finally points toward the empty
region, it switches back to its initial (positive) chiral state,
and the cycle repeats itself. For the particles in the dense region and
sufficiently away from the edge, the condition $P_i(d_c)>P_{\rm th}$ is
always satisfied, they all keep rotating clockwise on-site (their diffusivity is minimal)
and, therefore, appear to be brought together by one or more moving boundary
layers. This simple argument explains stationarity and orientation of the
edge currents for QS protocol 1 with appropriately restricted visual
aperture, $0.4 \lesssim \alpha/\pi \lesssim 0.8$ (see Sec. S4.B of Supplementary Text), their absence for QS
protocol 2, and implies that particles flowing along the boundary layers
change chirality sign with frequency of the order of $\omega_0$. All of this
is in good agreement with our simulation results (see movies in Supplementary Material).

A more quantitative analysis of the edge currents was performed by computing
the vorticity function $\varpi(r)$ [Fig. \ref{F2}(e)] (see Materials and Methods).
As expected, for $\omega_0>0$, the vorticity
is positive for clusters and negative for cavities. However, it is
non-zero only in a narrow annulus of radius $r \sim r_{c}$, which
delimits the structure boundary, consistently with the notion of edge
current. We also observed that the clusters' size, measured by $r_{c}$,
is proportional to $L$ and grows almost linearly with $d_c/L_p$ [Fig. \ref{F2}(e), inset],
in correspondence with the decrease of $\rho_c$ in Fig. \ref{F1}(i) (see Sec. S4.C of Supplementary Text for more details). Finally, by equating the
cluster diameter, $2 r_{c}$, to the box size, $L$, we get an estimate of the
$L$-independent $d_c$ value, when, in the two-phase configurations, clusters
get replaced by cavities [that is $d_c/L_p \gtrsim 1$ for the suspension
of Fig. \ref{F2}(e)].

The emergence of HU \cite{Torquato} was anticipated in
Fig. \ref{F1}(i): Upon increasing $d_c$, the high $\rho$ peak of the local
density distributions eventually disappears and the peak centered around
$\rho_0$ shrinks with respect to the corresponding (Gaussian) peak at very low/high
$d_c$. To better illustrate the HU properties of our system, we computed the
(time averaged) structure factor, $S(q)=(1/N)\sum_{i,j=1}^N \exp(i{\vec q}
\cdot [{\vec r}_i-{\vec r}_j])$, and density variance, $\langle \delta \rho^2
(l)\rangle = \langle [\rho(l) - \langle\rho(l)\rangle]^2\rangle$,  of the
stationary suspension in observation windows of size $l$ [Figs. \ref{F2} (f),(g)]. For
$1.3<d_c/L_p<1.6$,  we obtained clear-cut scaling laws, $S(q \to 0)\sim q^2$
[Fig. \ref{F2}(h)], and $\langle \delta \rho^2 (l \to \infty)\rangle \sim
l^{-3}$ [Fig. \ref{F2}(i)]. Consistently with the discussion in Ref.
\cite{NiSA}, we interpret these results as the signature of strong HU, the
closest to a perfect crystal structure. Within this $d_c$ range, HU spans
the entire simulation box. For smaller $d_c$ values, we analyzed the
particle distributions inside the two-phase structures, clusters and
cavities, alike. Restricting the computation of $S(q)$ and $\langle \delta
\rho^2 (l)\rangle$ to observation windows located inside the dense phase
[e.g., the large cluster and cavity of Figs. \ref{F2}(f),(g)] does not change the
exponents of the above scaling laws.
It should be noted that $S(q \to 0)$ is small but finite. As discussed in
Sec. S4.F of Supplementary Text (compare Figs. S10 and S11), this behavior arises mostly from the randomness inherent
to the visual QS function of Eq. (\ref{Pa}), and is therefore markedly suppressed
in suspension setups involving larger sensing ranges, $d_c$.
We consider this to be an example of {\it effective} HU \cite{NiSA}.

\vspace{5mm}
\noindent
\textbf{Chiral-passive mixtures}\\
To investigate the role of steric interactions in the present context, we simulated a colloidal
mixture containing only a small fraction, $\delta$, of {\em strongly} chiral
JP's. Active and passive disks have the same radius, and interact with each other
via the WCA potential. A weak translational noise term with $D_0\ll D_s$
(see Sec. S4.A of Supplementary Text) was incorporated to enhance the model's realism,
though it is not critical to the subsequent analysis.

\vspace{3mm}
\begin{figure*}
\centering \includegraphics[width=18cm]{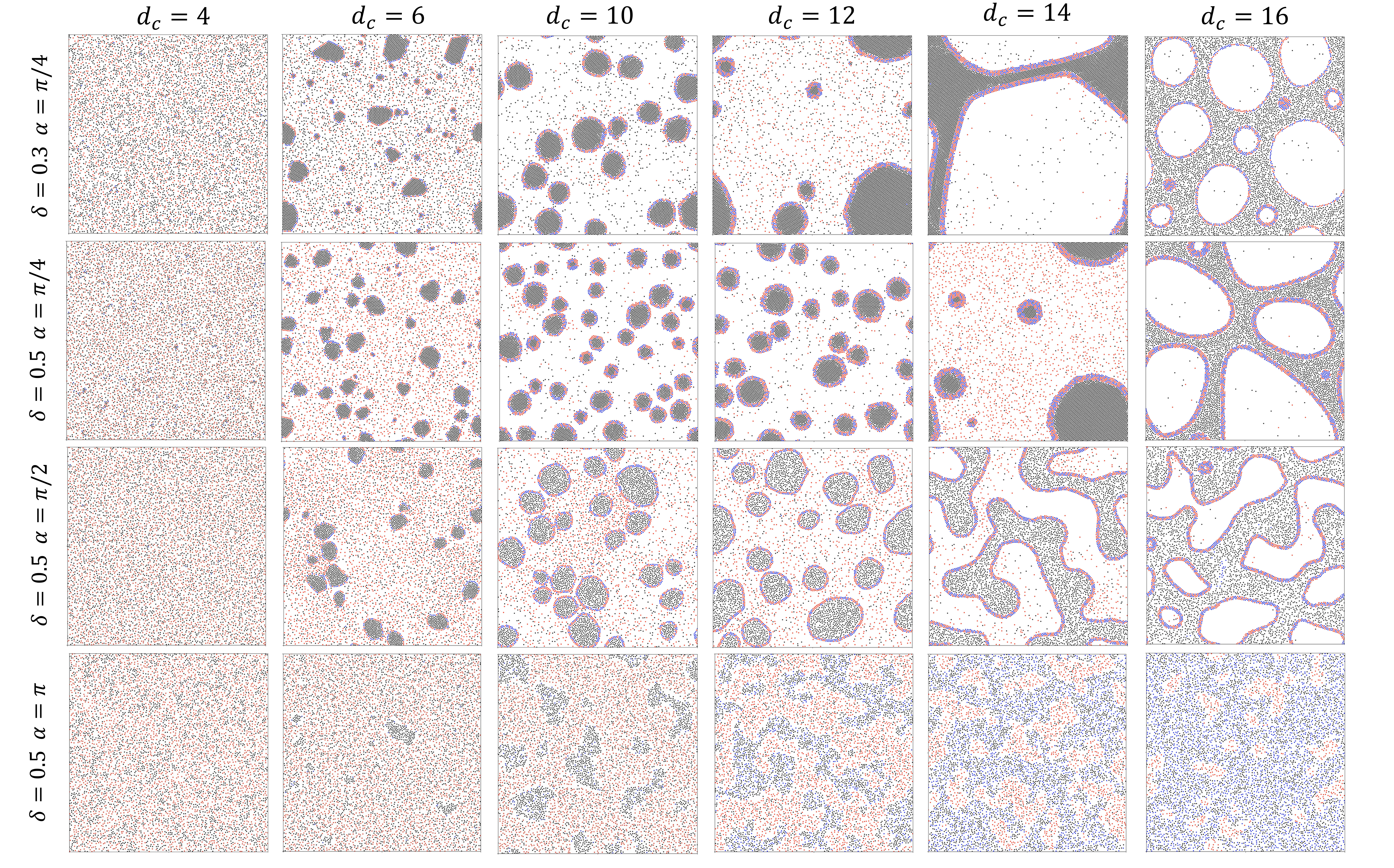}
\caption{Aggregation of passive particles by chiral particles under QS protocol 1 of Eq. (\ref{QS-1}) (implemented as explained in the text) for different $\delta$, $\alpha$, and $d_c$ (see legends) and in the presence of thermal fluctuation with $D_0=0.01$. $\delta$ is the fraction of active particles [blue (red) denoting positive (negative) chirality]; all the remaining (black) particles are passive. Disk parameters: $r_0=1$, $v_0=1$, and $D_\theta=0.01$; suspension parameters: $N=6\cdot 10^3$, $\bar \phi=0.21$, $\epsilon=1$; other QS protocol parameters: $L_p=10$, $\omega_0=1$.  Snapshots taken at $t=5\cdot 10^4$.
\label{F3}}
\end{figure*}

The QS threshold of Eq.(\ref{Pa}) should now be interpreted as follows: (i) QS applies exclusively
to active particles; (ii) the summation in Eq. (\ref{Pa}) is limited to
passive particles within the visual cone defined by parameters $\alpha$ and
$d_c$. Accordingly, the chirality of a tagged active particle remains unchanged even
in the presence of neighboring active particles, and the QS threshold
can be rewritten as $P_{\rm th}= (\alpha/\pi) \rho_p L_p$, where $\rho_p$
is the average density of the passive component in the mixture.

Our numerical findings for such a selective QS protocol are illustrated in Fig.
\ref{F3}. A striking observation is that at sufficiently large sensing distances,
the mixture develops spatial inhomogeneity, marked by the emergence of
multiple clusters. These structures feature a central core of passive
particles enveloped by one or more counterclockwise-rotating layers of active
particles. The stability and orientation of the edge currents align with the
mechanisms sketched in Fig. \ref{F1}(g). Notably, even a modest proportion of
active particles ($5\%$ or less, see Fig. S8) proves sufficient
to densely "herd'' passive colloids into compacted aggregates.

Key trends evident in Fig. \ref{F3} include:\\
{\em (i)} Increasing $d_c$ enlarges cluster size, causing the surrounding
dilute regions to contract and eventually fragment into predominantly empty
cavities encircled by clockwise-oriented edge currents. This behavior,
consistent with observations in the monodisperse suspensions of Figs. \ref{F1} and \ref{F2},
coincides with gradual melting of the cluster core. Concurrently, the
particle arrangement transitions sequentially from hexagonal crystalline to
hexatic and ultimately disordered.\\
{\em (ii)} Raising $\delta$ amplifies cluster multiplicity and accelerates
clustering onset at smaller $d_c$. This stems from the combined dynamics of a
greater number of visual QS disks, which reorganize more rapidly into rotating
boundary layers, resulting in smaller colloidal "herds." Additionally, at
fixed $d_c$, higher $\delta$ increases the number of active boundary layers
while reducing in-cluster particle density.\\
{\em (iii)} Smaller visual angles $\alpha$ are favored. Comparison of
snapshots in Fig. \ref{F3} reveals that for $\alpha=\pi/4$, clusters are more
abundant, structurally denser, and surrounded by thicker active shells
compared to $\alpha=\pi/2$. Consequently, the transition from clustered to
cavity configurations occurs at reduced $d_c$. Furthermore, simulations
conducted for $\alpha=\pi$ (representing the full visual cone or regular QS) reveal that
passive disks, as well as levo- and dextrogyre disks, tend to separate. In
agreement with the reasoning illustrated in Fig. \ref{F1}(g), regions
occupied by passive particles are not surrounded by edge currents.
The spatial arrangements of the mixture following this
dynamical separation display are depicted in Fig. \ref{F3} (bottom row).

\vspace{5mm}
\noindent
{\textcolor{myred}{\textbf{Discussion}}}\\
The most intriguing aspect of the collective dynamics in a visual QS chiral
suspension is that chirality changes of individual JPs, characterized
by a small gyration radius, can translate into large-scale dynamical effects
involving large portions of the suspension, namely edge currents and strong
HU. An additional effect of this mechanism is the enhancement of particle
diffusivity under HU conditions, well above its expected free-particle
value.  In Sec. S4.E, we also checked
the robustness of the reported effects against variations of the parameters
modeling the JP disk self-propulsion ($v_0$ and $D_\theta$), the QS protocol
($\alpha$ and $\omega_0$) and the pair potential ($\sigma$ and $\epsilon$).

Moreover, edge currents constitute a genuine QS
phenomenon which can arise even without steric interactions.
This conclusion is supported by numerical
simulations showing that dense structures surrounded by persistent edge
currents form spontaneously despite deactivating the WCA potential (i.e.,
setting the interaction strength $\epsilon=0$, see Sec. S4 of Supplementary Text).
Stated otherwise, the interfacial boundaries surrounding these condensed regions do not
generate confining pressure on the enclosed particles. The chirality-based
segregation of particles therefore originates not from excluded-volume
interactions, but rather emerges through a dynamical mechanism inherent the
adopted visual QS protocol. In contrast, herding of passive particles by QS chiral
particles occurs only in the presence of steric interactions between the two
particle species. Indeed, numerical simulation show that,
under the action of translational noise and in the absence of
steric repulsion, passive disks would traverse undisturbed
the encircling edge currents, thus restoring spatial uniformity.
\\

\vspace{5mm}
\noindent
{\textcolor{myred}{\textbf{Materials and Methods}}}\\
The suspension was assumed to be athermal. Accordingly,
in Eq. (\ref{LE}) we neglected thermal fluctuations against the angular
noise due to the self-propulsion mechanism \cite{ourPRL}. The
strength of the angular noise, $D_\theta$, defines the persistence (or correlation) time, $\tau_\theta=1/D_\theta$,
and length, $l_\theta=v_0/D_\theta$, of a free self-propelled
disk.  We remind that, for $t\gg \tau_\theta$, a free JP of fixed chiral frequency, $\omega_0$, would
undergo (non-Gaussian) normal diffusion
with diffusion constant $D_c=D_s/[1+(\omega_0 \tau_\theta)^2]$,
where $D_s=v_0^2/2D_\theta$ is the diffusion constant of
its achiral counterpart \cite{LoewenPRE78}. The stochastic differential Eqs. (\ref{LE}) were
numerically integrated by means of a standard Euler-Maruyama scheme \cite{Kloeden}.
To ensure numerical stability, the numerical integrations have been performed
using an appropriately short time step not larger than $\delta t=10^{-3}$.

The vorticity function displayed in Fig. \ref{F2}(e), is defined as
$\varpi(r)=[{1}/{N(r)}]\sum_{i=1}^{N(r)}{\bm \varphi_i}/{|{\bm \varphi}_i|}$,
where ${\bm \varphi_i}=({\vec r}_i-{\vec r}_{\times})\times {\vec v}_i/|{\vec
r}_i-{\vec r}_{\times}|^2$ and $N(r)$ is the number of particles with assigned
distance from the structure centroid, $r<|{\vec r}_i-{\vec r}_{\times}|<r+dr$. This
quantity was computed by choosing $N$ and $t$ so as to obtain single large
clusters or cavities, rounded in shape and with an easily identifiable centroid,
${\vec r}_c$ [see Fig. \ref{F2}(f),(g)].

Averages of $\varpi(r)$, $S(q)$ and $\delta \rho^2(l)$ where computed over hundreds
of steady-state configurations taken 200 time units apart.
[see caption of Fig. \ref{F2}].

The values of the length scales introduced in the Model section were chosen based on practical considerations:

(i) The disk radius, $r_0$ in the $V_{ij}$ potential, was assumed to be much smaller than the average inter-particle separation, $ l_L = (\pi r_0^2 / \bar{\phi})^{1/2} $ (quasi-pointlike particle regime). Note that edge currents are detectable also in the absence of steric repulsion.

(ii) The gyration radius, $ r_\omega = v_0 / \omega_0 $, was taken to be much smaller than the mean free path in a uniformly distributed achiral suspension, $ l_c = \pi r_0 / (2\bar{\phi}) $. This ensures that the chiral MIPS mechanism described in Ref.~\cite{NiSA} does not come into play, avoiding unnecessary complications.

(iii) To make the chiral nature of the active Janus disks experimentally detectable, we required $ \tau_\theta \omega_0 \gg 1 $, or equivalently, that $ r_\omega $ be much smaller than the persistence length, $ l_\theta = v_0 \tau_\theta $.

(iv) The QS protocol involves two more lengths, the sensing range, $d_c$, and the threshold defining range, $L_p$ [$P_{\rm th}$ in Eq. (\ref{QS-1})]. As shown in Fig.~\ref{F1}(f), the emergence of collective QS effects requires that the sensing range, $ d_c $, be comparable to or larger than $ l_L $.

(v) As a consequence of condition (iv), edge currents become detectable only when $ r_\omega \lesssim l_L $ (see Fig. S6). We note that below the MIPS threshold value of $ \bar{\phi} $ \cite{PRR7}, which is assumed throughout this work, it holds that $ l_c < l_L $. This imposes a tighter constraint on $ r_\omega $ (an upper bound) and correspondingly on $ \omega_0 $ (a lower bound), compared to condition (ii).

(vi) Finally, the box size $ L $ is not explicitly stated in the figure captions, as it is determined by our choice of the system size $ N $, the average packing fraction $ \bar{\phi} $, and the disk radius $ r_0 $.

$ r_0 $ and $ v_0 $, which are held constant throughout the simulations, simply define the units of length and time. Therefore, among all the length scales listed above, only $ r_\omega $, and $ l_L $ (i.e.,  $\omega_0$, $\bar \phi$) play a significant role in this study; $d_c$ (in units of $L_p$) quantifies the strength of the QS interaction. The other scales, $ l_c $ and $ l_\theta $, are set to be much larger than $ l_L $.

\vspace{5mm}
\noindent
{\textcolor{myred}{\textbf{References}}}

\makeatletter
\renewcommand{\bibsection}
\makeatother
\bibliography{arxiv}

\begin{thebibliography}{39}%
\makeatletter
\providecommand \@ifxundefined [1]{%
 \@ifx{#1\undefined}
}%
\providecommand \@ifnum [1]{%
 \ifnum #1\expandafter \@firstoftwo
 \else \expandafter \@secondoftwo
 \fi
}%
\providecommand \@ifx [1]{%
 \ifx #1\expandafter \@firstoftwo
 \else \expandafter \@secondoftwo
 \fi
}%
\providecommand \natexlab [1]{#1}%
\providecommand \enquote  [1]{``#1''}%
\providecommand \bibnamefont  [1]{#1}%
\providecommand \bibfnamefont [1]{#1}%
\providecommand \citenamefont [1]{#1}%
\providecommand \href@noop [0]{\@secondoftwo}%
\providecommand \href [0]{\begingroup \@sanitize@url \@href}%
\providecommand \@href[1]{\@@startlink{#1}\@@href}%
\providecommand \@@href[1]{\endgroup#1\@@endlink}%
\providecommand \@sanitize@url [0]{\catcode `\\12\catcode `\$12\catcode
  `\&12\catcode `\#12\catcode `\^12\catcode `\_12\catcode `\%12\relax}%
\providecommand \@@startlink[1]{}%
\providecommand \@@endlink[0]{}%
\providecommand \url  [0]{\begingroup\@sanitize@url \@url }%
\providecommand \@url [1]{\endgroup\@href {#1}{\urlprefix }}%
\providecommand \urlprefix  [0]{URL }%
\providecommand \Eprint [0]{\href }%
\providecommand \doibase [0]{https://doi.org/}%
\providecommand \selectlanguage [0]{\@gobble}%
\providecommand \bibinfo  [0]{\@secondoftwo}%
\providecommand \bibfield  [0]{\@secondoftwo}%
\providecommand \translation [1]{[#1]}%
\providecommand \BibitemOpen [0]{}%
\providecommand \bibitemStop [0]{}%
\providecommand \bibitemNoStop [0]{.\EOS\space}%
\providecommand \EOS [0]{\spacefactor3000\relax}%
\providecommand \BibitemShut  [1]{\csname bibitem#1\endcsname}%
\let\auto@bib@innerbib\@empty
\bibitem [{\citenamefont {Jiang}\ and\ \citenamefont
  {Granick}(2012)}]{Granick}%
  \BibitemOpen
  \bibfield  {author} {\bibinfo {author} {\bibfnamefont {S.}~\bibnamefont
  {Jiang}}\ and\ \bibinfo {author} {\bibfnamefont {S.}~\bibnamefont
  {Granick}},\ }\href@noop {} {\emph {\bibinfo {title} {Janus particle
  synthesis, self-assembly and applications}}}\ (\bibinfo  {publisher} {RSC
  Publishing, Cambridge},\ \bibinfo {year} {2012})\BibitemShut {NoStop}%
\bibitem [{\citenamefont {Walther}\ and\ \citenamefont
  {M\"{u}ller}(2013)}]{JP}%
  \BibitemOpen
  \bibfield  {author} {\bibinfo {author} {\bibfnamefont {A.}~\bibnamefont
  {Walther}}\ and\ \bibinfo {author} {\bibfnamefont {A.~H.~E.}\ \bibnamefont
  {M\"{u}ller}},\ }\bibfield  {title} {\bibinfo {title} {Janus particles:
  Synthesis, self-assembly, physical properties, and applications},\
  }\href@noop {} {\bibfield  {journal} {\bibinfo  {journal} {Chem. Rev.}\
  }\textbf {\bibinfo {volume} {113}},\ \bibinfo {pages} {5194} (\bibinfo {year}
  {2013})}\BibitemShut {NoStop}%
\bibitem [{\citenamefont {Marchetti}\ \emph {et~al.}(2013)\citenamefont
  {Marchetti}, \citenamefont {Joanny}, \citenamefont {Ramaswamy}, \citenamefont
  {Liverpool}, \citenamefont {Prost}, \citenamefont {Rao},\ and\ \citenamefont
  {Simha}}]{MarchettiRev}%
  \BibitemOpen
  \bibfield  {author} {\bibinfo {author} {\bibfnamefont {M.~C.}\ \bibnamefont
  {Marchetti}}, \bibinfo {author} {\bibfnamefont {J.~F.}\ \bibnamefont
  {Joanny}}, \bibinfo {author} {\bibfnamefont {S.}~\bibnamefont {Ramaswamy}},
  \bibinfo {author} {\bibfnamefont {T.~B.}\ \bibnamefont {Liverpool}}, \bibinfo
  {author} {\bibfnamefont {J.}~\bibnamefont {Prost}}, \bibinfo {author}
  {\bibfnamefont {M.}~\bibnamefont {Rao}},\ and\ \bibinfo {author}
  {\bibfnamefont {R.~A.}\ \bibnamefont {Simha}},\ }\bibfield  {title} {\bibinfo
  {title} {Hydrodynamics of soft active matter},\ }\href@noop {} {\bibfield
  {journal} {\bibinfo  {journal} {Rev. Mod. Phys.}\ }\textbf {\bibinfo {volume}
  {85}},\ \bibinfo {pages} {1143} (\bibinfo {year} {2013})}\BibitemShut
  {NoStop}%
\bibitem [{\citenamefont {Jayaram}\ \emph {et~al.}(2020)\citenamefont
  {Jayaram}, \citenamefont {Jie}, \citenamefont {Zhao},\ and\ \citenamefont
  {Andersson}}]{Andersson}%
  \BibitemOpen
  \bibfield  {author} {\bibinfo {author} {\bibfnamefont {R.}~\bibnamefont
  {Jayaram}}, \bibinfo {author} {\bibfnamefont {Y.}~\bibnamefont {Jie}},
  \bibinfo {author} {\bibfnamefont {L.}~\bibnamefont {Zhao}},\ and\ \bibinfo
  {author} {\bibfnamefont {H.~I.}\ \bibnamefont {Andersson}},\ }\bibfield
  {title} {\bibinfo {title} {Clustering of inertial spheres in evolving \
  {Taylor-Green} vortex flow},\ }\href@noop {} {\bibfield  {journal} {\bibinfo
  {journal} {Phys. Fluids}\ }\textbf {\bibinfo {volume} {32}},\ \bibinfo
  {pages} {043306} (\bibinfo {year} {2020})}\BibitemShut {NoStop}%
\bibitem [{\citenamefont {Li}\ \emph {et~al.}(2022)\citenamefont {Li},
  \citenamefont {Zhou}, \citenamefont {Marchesoni},\ and\ \citenamefont
  {Ghosh}}]{TU1}%
  \BibitemOpen
  \bibfield  {author} {\bibinfo {author} {\bibfnamefont {Y.}~\bibnamefont
  {Li}}, \bibinfo {author} {\bibfnamefont {Y.}~\bibnamefont {Zhou}}, \bibinfo
  {author} {\bibfnamefont {F.}~\bibnamefont {Marchesoni}},\ and\ \bibinfo
  {author} {\bibfnamefont {P.~K.}\ \bibnamefont {Ghosh}},\ }\bibfield  {title}
  {\bibinfo {title} {Colloidal clustering and diffusion in a convection cell
  array},\ }\href@noop {} {\bibfield  {journal} {\bibinfo  {journal} {Soft
  Matter}\ }\textbf {\bibinfo {volume} {18}},\ \bibinfo {pages} {4778}
  (\bibinfo {year} {2022})}\BibitemShut {NoStop}%
\bibitem [{\citenamefont {Ghosh}\ \emph {et~al.}(2023)\citenamefont {Ghosh},
  \citenamefont {Zhou}, \citenamefont {Li}, \citenamefont {Marchesoni},\ and\
  \citenamefont {Nori}}]{TU2}%
  \BibitemOpen
  \bibfield  {author} {\bibinfo {author} {\bibfnamefont {P.~K.}\ \bibnamefont
  {Ghosh}}, \bibinfo {author} {\bibfnamefont {Y.}~\bibnamefont {Zhou}},
  \bibinfo {author} {\bibfnamefont {Y.}~\bibnamefont {Li}}, \bibinfo {author}
  {\bibfnamefont {F.}~\bibnamefont {Marchesoni}},\ and\ \bibinfo {author}
  {\bibfnamefont {F.}~\bibnamefont {Nori}},\ }\bibfield  {title} {\bibinfo
  {title} {Binary mixtures in linear convection arrays},\ }\href@noop {}
  {\bibfield  {journal} {\bibinfo  {journal} {ChemPhysChem}\ }\textbf {\bibinfo
  {volume} {24}},\ \bibinfo {pages} {e202200471} (\bibinfo {year}
  {2023})}\BibitemShut {NoStop}%
\bibitem [{\citenamefont {Duan}\ \emph {et~al.}(2023)\citenamefont {Duan},
  \citenamefont {Agudo-Canalejo}, \citenamefont {Golestanian},\ and\
  \citenamefont {Mahault}}]{Golestanian}%
  \BibitemOpen
  \bibfield  {author} {\bibinfo {author} {\bibfnamefont {Y.}~\bibnamefont
  {Duan}}, \bibinfo {author} {\bibfnamefont {J.}~\bibnamefont
  {Agudo-Canalejo}}, \bibinfo {author} {\bibfnamefont {R.}~\bibnamefont
  {Golestanian}},\ and\ \bibinfo {author} {\bibfnamefont {B.}~\bibnamefont
  {Mahault}},\ }\bibfield  {title} {\bibinfo {title} {Dynamical pattern
  formation without self-attraction in quorum-sensing active matter: The
  interplay between nonreciprocity and motility},\ }\href@noop {} {\bibfield
  {journal} {\bibinfo  {journal} {Phys. Rev. Lett.}\ }\textbf {\bibinfo
  {volume} {131}},\ \bibinfo {pages} {148301} (\bibinfo {year}
  {2023})}\BibitemShut {NoStop}%
\bibitem [{\citenamefont {Fily}\ and\ \citenamefont {Marchetti}(2012)}]{Fily}%
  \BibitemOpen
  \bibfield  {author} {\bibinfo {author} {\bibfnamefont {Y.}~\bibnamefont
  {Fily}}\ and\ \bibinfo {author} {\bibfnamefont {M.~C.}\ \bibnamefont
  {Marchetti}},\ }\bibfield  {title} {\bibinfo {title} {Athermal phase
  separation of self-propelled particles with no alignment},\ }\href@noop {}
  {\bibfield  {journal} {\bibinfo  {journal} {Phys. Rev. Lett.}\ }\textbf
  {\bibinfo {volume} {108}},\ \bibinfo {pages} {235702} (\bibinfo {year}
  {2012})}\BibitemShut {NoStop}%
\bibitem [{\citenamefont {Redner}\ \emph
  {et~al.}(2013{\natexlab{a}})\citenamefont {Redner}, \citenamefont {Hagan},\
  and\ \citenamefont {Baskaran}}]{Redner}%
  \BibitemOpen
  \bibfield  {author} {\bibinfo {author} {\bibfnamefont {G.~S.}\ \bibnamefont
  {Redner}}, \bibinfo {author} {\bibfnamefont {M.~F.}\ \bibnamefont {Hagan}},\
  and\ \bibinfo {author} {\bibfnamefont {A.}~\bibnamefont {Baskaran}},\
  }\bibfield  {title} {\bibinfo {title} {Structure and dynamics of a
  phase-separating active colloidal fluid},\ }\href@noop {} {\bibfield
  {journal} {\bibinfo  {journal} {Phys. Rev. Lett.}\ }\textbf {\bibinfo
  {volume} {110}},\ \bibinfo {pages} {055701} (\bibinfo {year}
  {2013}{\natexlab{a}})}\BibitemShut {NoStop}%
\bibitem [{\citenamefont {Cates}\ and\ \citenamefont
  {Tailleur}(2015)}]{Cates2015}%
  \BibitemOpen
  \bibfield  {author} {\bibinfo {author} {\bibfnamefont {M.~E.}\ \bibnamefont
  {Cates}}\ and\ \bibinfo {author} {\bibfnamefont {J.}~\bibnamefont
  {Tailleur}},\ }\bibfield  {title} {\bibinfo {title} {Motility-induced phase
  separation},\ }\href@noop {} {\bibfield  {journal} {\bibinfo  {journal}
  {Annu. Rev. Condens. Matter Phys.}\ }\textbf {\bibinfo {volume} {6}},\
  \bibinfo {pages} {219} (\bibinfo {year} {2015})}\BibitemShut {NoStop}%
\bibitem [{\citenamefont {Miller}\ and\ \citenamefont {Bassler}(2001)}]{QS1}%
  \BibitemOpen
  \bibfield  {author} {\bibinfo {author} {\bibfnamefont {M.~B.}\ \bibnamefont
  {Miller}}\ and\ \bibinfo {author} {\bibfnamefont {B.~L.}\ \bibnamefont
  {Bassler}},\ }\bibfield  {title} {\bibinfo {title} {Quorum sensing in
  bacteria},\ }\href@noop {} {\bibfield  {journal} {\bibinfo  {journal} {Annu.
  Rev. Microbiol.}\ }\textbf {\bibinfo {volume} {55}},\ \bibinfo {pages} {165}
  (\bibinfo {year} {2001})}\BibitemShut {NoStop}%
\bibitem [{\citenamefont {Parsek}\ and\ \citenamefont {Greenberg}(2005)}]{QS2}%
  \BibitemOpen
  \bibfield  {author} {\bibinfo {author} {\bibfnamefont {M.~R.}\ \bibnamefont
  {Parsek}}\ and\ \bibinfo {author} {\bibfnamefont {E.~P.}\ \bibnamefont
  {Greenberg}},\ }\bibfield  {title} {\bibinfo {title} {Sociomicrobiology: the
  connections between quorum sensing and biofilms},\ }\href@noop {} {\bibfield
  {journal} {\bibinfo  {journal} {Trends Microbiol.}\ }\textbf {\bibinfo
  {volume} {13}},\ \bibinfo {pages} {27} (\bibinfo {year} {2005})}\BibitemShut
  {NoStop}%
\bibitem [{\citenamefont {B\"auerle}\ \emph {et~al.}(2018)\citenamefont
  {B\"auerle}, \citenamefont {Fischer}, \citenamefont {Speck},\ and\
  \citenamefont {Bechinger}}]{Bauerle2018}%
  \BibitemOpen
  \bibfield  {author} {\bibinfo {author} {\bibfnamefont {T.}~\bibnamefont
  {B\"auerle}}, \bibinfo {author} {\bibfnamefont {A.}~\bibnamefont {Fischer}},
  \bibinfo {author} {\bibfnamefont {T.}~\bibnamefont {Speck}},\ and\ \bibinfo
  {author} {\bibfnamefont {C.}~\bibnamefont {Bechinger}},\ }\bibfield  {title}
  {\bibinfo {title} {Self-organization of active particles by quorum sensing
  rules},\ }\href@noop {} {\bibfield  {journal} {\bibinfo  {journal} {Nat.
  Commun.}\ }\textbf {\bibinfo {volume} {9}},\ \bibinfo {pages} {3232}
  (\bibinfo {year} {2018})}\BibitemShut {NoStop}%
\bibitem [{\citenamefont {Lavergne}\ \emph {et~al.}(2019)\citenamefont
  {Lavergne}, \citenamefont {Wendehenne}, \citenamefont {B\"auerle},\ and\
  \citenamefont {Bechinger}}]{Lavergne2019}%
  \BibitemOpen
  \bibfield  {author} {\bibinfo {author} {\bibfnamefont {F.~A.}\ \bibnamefont
  {Lavergne}}, \bibinfo {author} {\bibfnamefont {H.}~\bibnamefont
  {Wendehenne}}, \bibinfo {author} {\bibfnamefont {T.}~\bibnamefont
  {B\"auerle}},\ and\ \bibinfo {author} {\bibfnamefont {C.}~\bibnamefont
  {Bechinger}},\ }\bibfield  {title} {\bibinfo {title} {Group formation and
  cohesion of active particles with visual perception–dependent motility},\
  }\href@noop {} {\bibfield  {journal} {\bibinfo  {journal} {Science}\ }\textbf
  {\bibinfo {volume} {364}},\ \bibinfo {pages} {6435} (\bibinfo {year}
  {2019})}\BibitemShut {NoStop}%
\bibitem [{\citenamefont {Barberis}\ and\ \citenamefont
  {Peruani}(2016)}]{Peruani}%
  \BibitemOpen
  \bibfield  {author} {\bibinfo {author} {\bibfnamefont {L.}~\bibnamefont
  {Barberis}}\ and\ \bibinfo {author} {\bibfnamefont {F.}~\bibnamefont
  {Peruani}},\ }\bibfield  {title} {\bibinfo {title} {Large-scale patterns in a
  minimal cognitive flocking model: Incidental leaders, nematic patterns, and
  aggregates},\ }\href@noop {} {\bibfield  {journal} {\bibinfo  {journal}
  {Phys. Rev. Lett.}\ }\textbf {\bibinfo {volume} {117}},\ \bibinfo {pages}
  {248001} (\bibinfo {year} {2016})}\BibitemShut {NoStop}%
\bibitem [{\citenamefont {B\"auerle}\ \emph {et~al.}(2020)\citenamefont
  {B\"auerle}, \citenamefont {R.~C},\ and\ \citenamefont
  {Bechinger}}]{Bauerle2020}%
  \BibitemOpen
  \bibfield  {author} {\bibinfo {author} {\bibfnamefont {T.}~\bibnamefont
  {B\"auerle}}, \bibinfo {author} {\bibfnamefont {L.}~\bibnamefont {R.~C}},\
  and\ \bibinfo {author} {\bibfnamefont {C.}~\bibnamefont {Bechinger}},\
  }\bibfield  {title} {\bibinfo {title} {Formation of stable and responsive
  collective states in suspensions of active colloids},\ }\href@noop {}
  {\bibfield  {journal} {\bibinfo  {journal} {Nat. Commun.}\ }\textbf {\bibinfo
  {volume} {11}},\ \bibinfo {pages} {2547} (\bibinfo {year}
  {2020})}\BibitemShut {NoStop}%
\bibitem [{\citenamefont {Thapa}\ \emph {et~al.}(2024)\citenamefont {Thapa},
  \citenamefont {Pinchasik},\ and\ \citenamefont {Shokef}}]{Israel}%
  \BibitemOpen
  \bibfield  {author} {\bibinfo {author} {\bibfnamefont {S.}~\bibnamefont
  {Thapa}}, \bibinfo {author} {\bibfnamefont {B.-E.}\ \bibnamefont
  {Pinchasik}},\ and\ \bibinfo {author} {\bibfnamefont {Y.}~\bibnamefont
  {Shokef}},\ }\bibfield  {title} {\bibinfo {title} {Emergent clustering due to
  quorum sensing in active matter},\ }\href@noop {} {\bibfield  {journal}
  {\bibinfo  {journal} {New J. Phys.}\ }\textbf {\bibinfo {volume} {26}},\
  \bibinfo {pages} {023010} (\bibinfo {year} {2024})}\BibitemShut {NoStop}%
\bibitem [{\citenamefont {Zhou}\ \emph {et~al.}(2023)\citenamefont {Zhou},
  \citenamefont {Li},\ and\ \citenamefont {Marchesoni}}]{ourCPL}%
  \BibitemOpen
  \bibfield  {author} {\bibinfo {author} {\bibfnamefont {Y.}~\bibnamefont
  {Zhou}}, \bibinfo {author} {\bibfnamefont {Y.}~\bibnamefont {Li}},\ and\
  \bibinfo {author} {\bibfnamefont {F.}~\bibnamefont {Marchesoni}},\ }\bibfield
   {title} {\bibinfo {title} {A quorum sensing active matter in a confined
  geometry},\ }\href@noop {} {\bibfield  {journal} {\bibinfo  {journal} {Chin.
  Phys. Lett.}\ }\textbf {\bibinfo {volume} {40}},\ \bibinfo {pages} {100505}
  (\bibinfo {year} {2023})}\BibitemShut {NoStop}%
\bibitem [{\citenamefont {Saavedra}\ \emph {et~al.}(2024)\citenamefont
  {Saavedra}, \citenamefont {Gompper},\ and\ \citenamefont {Ripoll}}]{Ripoll}%
  \BibitemOpen
  \bibfield  {author} {\bibinfo {author} {\bibfnamefont {R.}~\bibnamefont
  {Saavedra}}, \bibinfo {author} {\bibfnamefont {G.}~\bibnamefont {Gompper}},\
  and\ \bibinfo {author} {\bibfnamefont {M.}~\bibnamefont {Ripoll}},\
  }\bibfield  {title} {\bibinfo {title} {Swirling due to misaligned
  perception-dependent motility},\ }\href@noop {} {\bibfield  {journal}
  {\bibinfo  {journal} {Phys. Rev. Lett.}\ }\textbf {\bibinfo {volume} {132}},\
  \bibinfo {pages} {268301} (\bibinfo {year} {2024})}\BibitemShut {NoStop}%
\bibitem [{\citenamefont {Negi}\ \emph {et~al.}(2022)\citenamefont {Negi},
  \citenamefont {Winkler},\ and\ \citenamefont {Gompper}}]{Gompper}%
  \BibitemOpen
  \bibfield  {author} {\bibinfo {author} {\bibfnamefont {R.~S.}\ \bibnamefont
  {Negi}}, \bibinfo {author} {\bibfnamefont {R.~G.}\ \bibnamefont {Winkler}},\
  and\ \bibinfo {author} {\bibfnamefont {G.}~\bibnamefont {Gompper}},\
  }\bibfield  {title} {\bibinfo {title} {Emergent collective behavior of active
  brownian particles with visual perception},\ }\href@noop {} {\bibfield
  {journal} {\bibinfo  {journal} {Soft Matter}\ }\textbf {\bibinfo {volume}
  {18}},\ \bibinfo {pages} {6167} (\bibinfo {year} {2022})}\BibitemShut
  {NoStop}%
\bibitem [{\citenamefont {Lei}\ \emph {et~al.}(2019)\citenamefont {Lei},
  \citenamefont {Ciamarra},\ and\ \citenamefont {Ni}}]{NiSA}%
  \BibitemOpen
  \bibfield  {author} {\bibinfo {author} {\bibfnamefont {Q.}~\bibnamefont
  {Lei}}, \bibinfo {author} {\bibfnamefont {M.~P.}\ \bibnamefont {Ciamarra}},\
  and\ \bibinfo {author} {\bibfnamefont {R.}~\bibnamefont {Ni}},\ }\bibfield
  {title} {\bibinfo {title} {Nonequilibrium strongly hyperuniform fluids of
  circle active particles with large local density fluctuations},\ }\href@noop
  {} {\bibfield  {journal} {\bibinfo  {journal} {Sci. Adv.}\ }\textbf {\bibinfo
  {volume} {5}},\ \bibinfo {pages} {eaau7423} (\bibinfo {year}
  {2019})}\BibitemShut {NoStop}%
\bibitem [{\citenamefont {Lauga}\ and\ \citenamefont {Powers}(2009)}]{Lauga}%
  \BibitemOpen
  \bibfield  {author} {\bibinfo {author} {\bibfnamefont {E.}~\bibnamefont
  {Lauga}}\ and\ \bibinfo {author} {\bibfnamefont {T.~R.}\ \bibnamefont
  {Powers}},\ }\bibfield  {title} {\bibinfo {title} {The hydrodynamics of
  swimming microorganisms},\ }\href@noop {} {\bibfield  {journal} {\bibinfo
  {journal} {Rep. Prog. Phys.}\ }\textbf {\bibinfo {volume} {72}},\ \bibinfo
  {pages} {096601} (\bibinfo {year} {2009})}\BibitemShut {NoStop}%
\bibitem [{\citenamefont {Liao}\ and\ \citenamefont {Klapp}(2018)}]{Klapp}%
  \BibitemOpen
  \bibfield  {author} {\bibinfo {author} {\bibfnamefont {G.-J.}\ \bibnamefont
  {Liao}}\ and\ \bibinfo {author} {\bibfnamefont {S.~H.~L.}\ \bibnamefont
  {Klapp}},\ }\bibfield  {title} {\bibinfo {title} {Clustering and phase
  separation of circle swimmers dispersed in a monolayer},\ }\href@noop {}
  {\bibfield  {journal} {\bibinfo  {journal} {Soft Matter}\ }\textbf {\bibinfo
  {volume} {14}},\ \bibinfo {pages} {7873} (\bibinfo {year}
  {2018})}\BibitemShut {NoStop}%
\bibitem [{\citenamefont {Ma}\ and\ \citenamefont {Ni}(2022)}]{NiJCP}%
  \BibitemOpen
  \bibfield  {author} {\bibinfo {author} {\bibfnamefont {Z.}~\bibnamefont
  {Ma}}\ and\ \bibinfo {author} {\bibfnamefont {R.}~\bibnamefont {Ni}},\
  }\bibfield  {title} {\bibinfo {title} {Dynamical clustering interrupts
  motility-induced phase separation in chiral active brownian particles},\
  }\href@noop {} {\bibfield  {journal} {\bibinfo  {journal} {J. Chem. Phys.}\
  }\textbf {\bibinfo {volume} {156}},\ \bibinfo {pages} {021102} (\bibinfo
  {year} {2022})}\BibitemShut {NoStop}%
\bibitem [{\citenamefont {Lei}\ and\ \citenamefont {Ni}(2019)}]{NiPNAS}%
  \BibitemOpen
  \bibfield  {author} {\bibinfo {author} {\bibfnamefont {Q.}~\bibnamefont
  {Lei}}\ and\ \bibinfo {author} {\bibfnamefont {R.}~\bibnamefont {Ni}},\
  }\bibfield  {title} {\bibinfo {title} {Hydrodynamics of random-organizing
  hyperuniform fluids},\ }\href@noop {} {\bibfield  {journal} {\bibinfo
  {journal} {Proc. Natl. Acad. Sci. U. S. A.}\ }\textbf {\bibinfo {volume}
  {116}},\ \bibinfo {pages} {22983} (\bibinfo {year} {2019})}\BibitemShut
  {NoStop}%
\bibitem [{\citenamefont {Torquato}\ and\ \citenamefont
  {Stillinger}(2003)}]{Stillinger}%
  \BibitemOpen
  \bibfield  {author} {\bibinfo {author} {\bibfnamefont {S.}~\bibnamefont
  {Torquato}}\ and\ \bibinfo {author} {\bibfnamefont {F.~H.}\ \bibnamefont
  {Stillinger}},\ }\bibfield  {title} {\bibinfo {title} {Local density
  fluctuations, hyperuniformity, and order metrics},\ }\href@noop {} {\bibfield
   {journal} {\bibinfo  {journal} {Phys. Rev. E}\ }\textbf {\bibinfo {volume}
  {68}},\ \bibinfo {pages} {041113} (\bibinfo {year} {2003})}\BibitemShut
  {NoStop}%
\bibitem [{\citenamefont {Torquato}(2018)}]{Torquato}%
  \BibitemOpen
  \bibfield  {author} {\bibinfo {author} {\bibfnamefont {S.}~\bibnamefont
  {Torquato}},\ }\bibfield  {title} {\bibinfo {title} {Hyperuniform states of
  matter},\ }\href@noop {} {\bibfield  {journal} {\bibinfo  {journal} {Phys.
  Rep}\ }\textbf {\bibinfo {volume} {745}},\ \bibinfo {pages} {1} (\bibinfo
  {year} {2018})}\BibitemShut {NoStop}%
\bibitem [{\citenamefont {Weeks}\ \emph {et~al.}(1971)\citenamefont {Weeks},
  \citenamefont {Chandler},\ and\ \citenamefont {Andersen}}]{CWA}%
  \BibitemOpen
  \bibfield  {author} {\bibinfo {author} {\bibfnamefont {J.~D.}\ \bibnamefont
  {Weeks}}, \bibinfo {author} {\bibfnamefont {D.}~\bibnamefont {Chandler}},\
  and\ \bibinfo {author} {\bibfnamefont {H.~C.}\ \bibnamefont {Andersen}},\
  }\bibfield  {title} {\bibinfo {title} {Role of repulsive forces in
  determining the equilibrium structure of simple liquids},\ }\href@noop {}
  {\bibfield  {journal} {\bibinfo  {journal} {J. Chem. Phys.}\ }\textbf
  {\bibinfo {volume} {54}},\ \bibinfo {pages} {5237} (\bibinfo {year}
  {1971})}\BibitemShut {NoStop}%
\bibitem [{\citenamefont {Yang}\ \emph {et~al.}(2017)\citenamefont {Yang},
  \citenamefont {Liu}, \citenamefont {Li}, \citenamefont {Marchesoni},
  \citenamefont {H\"anggi},\ and\ \citenamefont {Zhang}}]{PNAS}%
  \BibitemOpen
  \bibfield  {author} {\bibinfo {author} {\bibfnamefont {X.}~\bibnamefont
  {Yang}}, \bibinfo {author} {\bibfnamefont {C.}~\bibnamefont {Liu}}, \bibinfo
  {author} {\bibfnamefont {Y.}~\bibnamefont {Li}}, \bibinfo {author}
  {\bibfnamefont {F.}~\bibnamefont {Marchesoni}}, \bibinfo {author}
  {\bibfnamefont {P.}~\bibnamefont {H\"anggi}},\ and\ \bibinfo {author}
  {\bibfnamefont {H.~P.}\ \bibnamefont {Zhang}},\ }\bibfield  {title} {\bibinfo
  {title} {Hydrodynamic and entropic effects on colloidal diffusion in
  corrugated channels},\ }\href@noop {} {\bibfield  {journal} {\bibinfo
  {journal} {Proc. Natl. Acad. Sci. U.S.A.}\ }\textbf {\bibinfo {volume}
  {114}},\ \bibinfo {pages} {9564} (\bibinfo {year} {2017})}\BibitemShut
  {NoStop}%
\bibitem [{\citenamefont {Takagi}\ \emph {et~al.}(2014)\citenamefont {Takagi},
  \citenamefont {Palacci}, \citenamefont {Braunschweig}, \citenamefont
  {Shelley},\ and\ \citenamefont {Zhang}}]{Takagi}%
  \BibitemOpen
  \bibfield  {author} {\bibinfo {author} {\bibfnamefont {D.}~\bibnamefont
  {Takagi}}, \bibinfo {author} {\bibfnamefont {J.}~\bibnamefont {Palacci}},
  \bibinfo {author} {\bibfnamefont {A.~B.}\ \bibnamefont {Braunschweig}},
  \bibinfo {author} {\bibfnamefont {M.~J.}\ \bibnamefont {Shelley}},\ and\
  \bibinfo {author} {\bibfnamefont {J.}~\bibnamefont {Zhang}},\ }\bibfield
  {title} {\bibinfo {title} {Hydrodynamic capture of microswimmers into
  sphere-bound orbits},\ }\href@noop {} {\bibfield  {journal} {\bibinfo
  {journal} {Soft Matter}\ }\textbf {\bibinfo {volume} {10}},\ \bibinfo {pages}
  {1784} (\bibinfo {year} {2014})}\BibitemShut {NoStop}%
\bibitem [{\citenamefont {Berlinger}\ \emph {et~al.}(2021)\citenamefont
  {Berlinger}, \citenamefont {Gauci},\ and\ \citenamefont {Nagpal}}]{fish1}%
  \BibitemOpen
  \bibfield  {author} {\bibinfo {author} {\bibfnamefont {F.}~\bibnamefont
  {Berlinger}}, \bibinfo {author} {\bibfnamefont {M.}~\bibnamefont {Gauci}},\
  and\ \bibinfo {author} {\bibfnamefont {R.}~\bibnamefont {Nagpal}},\
  }\bibfield  {title} {\bibinfo {title} {Implicit coordination for 3d
  underwater collective behaviors in a fish-inspired robot swarm},\ }\href@noop
  {} {\bibfield  {journal} {\bibinfo  {journal} {Sci Robot.}\ }\textbf
  {\bibinfo {volume} {6}},\ \bibinfo {pages} {eabd8668} (\bibinfo {year}
  {2021})}\BibitemShut {NoStop}%
\bibitem [{\citenamefont {Jiang}\ \emph {et~al.}(2023)\citenamefont {Jiang},
  \citenamefont {Zhou}, \citenamefont {Chen}, \citenamefont {Yang},
  \citenamefont {Dong},\ and\ \citenamefont {Wang}}]{fish2}%
  \BibitemOpen
  \bibfield  {author} {\bibinfo {author} {\bibfnamefont {M.}~\bibnamefont
  {Jiang}}, \bibinfo {author} {\bibfnamefont {A.}~\bibnamefont {Zhou}},
  \bibinfo {author} {\bibfnamefont {R.}~\bibnamefont {Chen}}, \bibinfo {author}
  {\bibfnamefont {Y.}~\bibnamefont {Yang}}, \bibinfo {author} {\bibfnamefont
  {H.}~\bibnamefont {Dong}},\ and\ \bibinfo {author} {\bibfnamefont
  {W.}~\bibnamefont {Wang}},\ }\bibfield  {title} {\bibinfo {title} {Collective
  motions of fish originate from balanced local perceptual interactions and
  individual stochastic},\ }\href@noop {} {\bibfield  {journal} {\bibinfo
  {journal} {Phys. Rev. E}\ }\textbf {\bibinfo {volume} {107}},\ \bibinfo
  {pages} {024411} (\bibinfo {year} {2023})}\BibitemShut {NoStop}%
\bibitem [{\citenamefont {Zhou}\ \emph {et~al.}(2024)\citenamefont {Zhou},
  \citenamefont {Li},\ and\ \citenamefont {Marchesoni}}]{NSO}%
  \BibitemOpen
  \bibfield  {author} {\bibinfo {author} {\bibfnamefont {Y.}~\bibnamefont
  {Zhou}}, \bibinfo {author} {\bibfnamefont {Y.}~\bibnamefont {Li}},\ and\
  \bibinfo {author} {\bibfnamefont {F.}~\bibnamefont {Marchesoni}},\ }\bibfield
   {title} {\bibinfo {title} {Clustering of quorum sensing colloidal
  particles},\ }\href@noop {} {\bibfield  {journal} {\bibinfo  {journal} {Natl.
  Sci. Open}\ }\textbf {\bibinfo {volume} {3}},\ \bibinfo {pages} {20230081}
  (\bibinfo {year} {2024})}\BibitemShut {NoStop}%
\bibitem [{\citenamefont {Okabe}\ and\ \citenamefont
  {Sugihara}(2000)}]{Voronoi}%
  \BibitemOpen
  \bibfield  {author} {\bibinfo {author} {\bibfnamefont {A.}~\bibnamefont
  {Okabe}}\ and\ \bibinfo {author} {\bibfnamefont {K.}~\bibnamefont
  {Sugihara}},\ }\href@noop {} {\emph {\bibinfo {title} {Spatial Tessellations:
  Concepts and Applications of Voronoi Diagrams}}}\ (\bibinfo  {publisher} {2nd
  ed. (Wiley, New York)},\ \bibinfo {year} {2000})\BibitemShut {NoStop}%
\bibitem [{\citenamefont {Redner}\ \emph
  {et~al.}(2013{\natexlab{b}})\citenamefont {Redner}, \citenamefont {Hagan},\
  and\ \citenamefont {Baskaran}}]{Redner1}%
  \BibitemOpen
  \bibfield  {author} {\bibinfo {author} {\bibfnamefont {G.~S.}\ \bibnamefont
  {Redner}}, \bibinfo {author} {\bibfnamefont {M.~F.}\ \bibnamefont {Hagan}},\
  and\ \bibinfo {author} {\bibfnamefont {A.}~\bibnamefont {Baskaran}},\
  }\bibfield  {title} {\bibinfo {title} {Structure and dynamics of a
  phase-separating active colloidal fluid},\ }\href@noop {} {\bibfield
  {journal} {\bibinfo  {journal} {Phys. Rev. Lett.}\ }\textbf {\bibinfo
  {volume} {110}},\ \bibinfo {pages} {055701} (\bibinfo {year}
  {2013}{\natexlab{b}})}\BibitemShut {NoStop}%
\bibitem [{\citenamefont {Nayak}\ \emph {et~al.}(2025)\citenamefont {Nayak},
  \citenamefont {Bag}, \citenamefont {Ghosh}, \citenamefont {Li}, \citenamefont
  {Zhou}, \citenamefont {Yin}, \citenamefont {Marchesoni},\ and\ \citenamefont
  {Nori}}]{PRR7}%
  \BibitemOpen
  \bibfield  {author} {\bibinfo {author} {\bibfnamefont {S.}~\bibnamefont
  {Nayak}}, \bibinfo {author} {\bibfnamefont {P.}~\bibnamefont {Bag}}, \bibinfo
  {author} {\bibfnamefont {P.~K.}\ \bibnamefont {Ghosh}}, \bibinfo {author}
  {\bibfnamefont {Y.}~\bibnamefont {Li}}, \bibinfo {author} {\bibfnamefont
  {Y.}~\bibnamefont {Zhou}}, \bibinfo {author} {\bibfnamefont {Q.}~\bibnamefont
  {Yin}}, \bibinfo {author} {\bibfnamefont {F.}~\bibnamefont {Marchesoni}},\
  and\ \bibinfo {author} {\bibfnamefont {F.}~\bibnamefont {Nori}},\ }\bibfield
  {title} {\bibinfo {title} {Diffusion transients in motility-induced phase
  separation},\ }\href@noop {} {\bibfield  {journal} {\bibinfo  {journal}
  {Phys. Rev. Res.}\ }\textbf {\bibinfo {volume} {7}},\ \bibinfo {pages}
  {013153} (\bibinfo {year} {2025})}\BibitemShut {NoStop}%
\bibitem [{\citenamefont {Ghosh}\ \emph {et~al.}(2013)\citenamefont {Ghosh},
  \citenamefont {Misko}, \citenamefont {Marchesoni},\ and\ \citenamefont
  {Nori}}]{ourPRL}%
  \BibitemOpen
  \bibfield  {author} {\bibinfo {author} {\bibfnamefont {P.~K.}\ \bibnamefont
  {Ghosh}}, \bibinfo {author} {\bibfnamefont {V.~R.}\ \bibnamefont {Misko}},
  \bibinfo {author} {\bibfnamefont {F.}~\bibnamefont {Marchesoni}},\ and\
  \bibinfo {author} {\bibfnamefont {F.}~\bibnamefont {Nori}},\ }\bibfield
  {title} {\bibinfo {title} {Self-propelled janus particles in a ratchet:
  Numerical simulations},\ }\href@noop {} {\bibfield  {journal} {\bibinfo
  {journal} {Phys. Rev. Lett.}\ }\textbf {\bibinfo {volume} {110}},\ \bibinfo
  {pages} {268301} (\bibinfo {year} {2013})}\BibitemShut {NoStop}%
\bibitem [{\citenamefont {van Teeffelen}\ and\ \citenamefont
  {L\"owen}(2008)}]{LoewenPRE78}%
  \BibitemOpen
  \bibfield  {author} {\bibinfo {author} {\bibfnamefont {S.}~\bibnamefont {van
  Teeffelen}}\ and\ \bibinfo {author} {\bibfnamefont {H.}~\bibnamefont
  {L\"owen}},\ }\bibfield  {title} {\bibinfo {title} {Dynamics of a brownian
  circle swimmer},\ }\href@noop {} {\bibfield  {journal} {\bibinfo  {journal}
  {Phys. Rev. E}\ }\textbf {\bibinfo {volume} {78}},\ \bibinfo {pages} {020101}
  (\bibinfo {year} {2008})}\BibitemShut {NoStop}%
\bibitem [{\citenamefont {Kloeden}\ and\ \citenamefont
  {Platen}(1992)}]{Kloeden}%
  \BibitemOpen
  \bibfield  {author} {\bibinfo {author} {\bibfnamefont {P.~E.}\ \bibnamefont
  {Kloeden}}\ and\ \bibinfo {author} {\bibfnamefont {E.}~\bibnamefont
  {Platen}},\ }\href@noop {} {\emph {\bibinfo {title} {Numerical Solution of
  Stochastic Differential Equations}}}\ (\bibinfo  {publisher} {Springer,
  Berlin},\ \bibinfo {year} {1992})\BibitemShut {NoStop}%
\end{thebibliography}%

\vspace{5mm}
\noindent
{\textcolor{myred}{\textbf{Acknowledgements}}}\\
\textbf{Funding:} This work is supported by National Natural Science Foundation of China grant 12375037 (LYY), National Natural Science Foundation of China grant 12350710786 (MF, LYY), SERB Core Research Grant CRG/2021/007394 (GPK), CSIR EMR II file no. 01/3115/23 (GPK), UGC, New Delhi, India Senior Research Fellowship (BP). \textbf {Author contributions: } LYY, GPK and MF designed the research; ZY, YQ, NS and BP performed research; All authors contributed to the data analysis; MF wrote the manuscript; GPK and LYY edited it.  \textbf{Competing interests:} The authors declare that they have no competing interests. \textbf{Data and materials availability:} All data needed to evaluate the conclusions in the paper are presented in the paper and/or Supplementary Materials. Additional data related to this paper may be requested from the authors.

\end{document}